\documentclass[11pt,letterpaper]{article}
\usepackage{jheppub}
\usepackage{pstool}
\usepackage{amsmath}
\usepackage{amssymb}
\usepackage[config=altsf]{subfig}
\usepackage{dsfont}
\usepackage{array}
\usepackage{slashed}
\usepackage{afterpage}
\usepackage{appendix}
\usepackage{multirow}

\usepackage{feynmp}
\makeatletter
\def\endfmffile{%
  \fmfcmd{\p@rcent\space the end.^^J%
          end.^^J%
          endinput;}%
  \if@fmfio
    \immediate\closeout\@outfmf
  \fi
  \ifnum\pdfshellescape=\@ne
    \immediate\write18{mpost \thefmffile}%
  \fi}
\makeatother

\title{Mixing stops at the LHC}
\author{Prateek Agrawal}
\author{and Claudia Frugiuele}
\affiliation{Fermilab, P.O. Box 500, Batavia, IL 60510, USA}
\emailAdd{prateek@fnal.gov}
\emailAdd{claudiaf@fnal.gov}

\abstract { 
We study the phenomenology of a light stop NLSP 
in the presence of large mixing with either the first or the second
generation.
R-symmetric models provide a prime setting for this scenario, but our
discussion also applies to the MSSM when a
significant amount of mixing can be accommodated.  In our framework the
dominant stop decay is through the flavor violating mode into a light
jet and the LSP in an extended region of parameter space.
There are currently no limits  from ATLAS
and CMS in this region.  We emulate shape-based hadronic SUSY searches
for this topology, and find that they have potential sensitivity. 
If the extension of these analyses to this region  is robust,  we find that these searches can set
strong exclusion limits on light stops. If not, then the flavor
violating decay mode is challenging and may represent a blind spot in
stop searches even at 13 TeV. Thus, an experimental investigation of this scenario is well motivated.  }

\preprint{FERMILAB-PUB-13-095-T}
\begin{document}
\maketitle

\section{Introduction} 
With the discovery of the particle consistent with standard model (SM) Higgs
boson \cite{Aad:2012tfa,Chatrchyan:2012ufa}, the hierarchy problem
takes on a concrete
and immediate nature.
%
Low energy supersymmetry  is one of the most studied and robust
solutions to the hierarchy problem.  A simple realization of
supersymmetry is the Minimal Supersymmetric Standard Model (MSSM),
which in its most general form suffers from a very acute flavor
problem \cite{Dimopoulos:1995ju,Gabbiani:1996hi}.  The tension between
flavor and natural low energy SUSY can be economically solved in models which predict
spectra with degenerate squarks (and sleptons).
In the pre-LHC era this paradigm provided a compelling framework of
flavor in low energy SUSY models.  However, the LHC bounds on
these models push the spectrum of new states 
to the TeV scale, already
requiring the SM to suffer some amount of fine-tuning. 

Natural SUSY models surviving the LHC bounds now require a non-trivial flavor structure for the squark mass
matrix so as to produce a hierarchical spectrum with the first and
the second generation significantly heavier than the third generation
\cite{Papucci:2011wy,Brust:2011tb}.  In the MSSM  the structure of
possible squark mass matrices is severely restricted by low energy
observables
\cite{Giudice:2008uk,Barbieri:2010ar,Barbieri:2010pd,Craig:2012di}.
R-symmetric models 
provide a compelling setting for investigating
more complicated flavor structures since they generically admit larger
mixing among the squarks than in the MSSM \cite{Kribs:2007ac}. 
This is a consequence of
the fact that
most flavor violating observables turn out to be proportional to terms
which break R-symmetry, and are hence  suppressed \cite{Kribs:2007ac,
Fok:2010vk,Blechman:2008gu}.  Therefore, a
significant amount of flavor mixing is possible and quite generic.
This could be an interesting ingredient in building
flavorful SUSY breaking/mediation models, and also
significantly alter the LHC signatures of SUSY particles
\cite{Kribs:2009zy}. 

Models with an (approximate) R-symmetry  \cite{Kribs:2007ac, Frugiuele:2011mh,Davies:2011mp,Frugiuele:2012pe,Riva:2012hz}
are additionally interesting as an alternative to the MSSM since they
require the gauginos to be Dirac fermions.\footnote{ The R-symmetry cannot be an exact symmetry since it is at
least broken by the gravitino mass. The effects of the R-breaking
are therefore communicated to the visible sector through anomaly
mediation, see \cite{Kribs:2007ac}.}
This makes them less minimal than the MSSM in terms of the particle content,
 adding adjoint superfields for each SM gauge group, but on the other hand
 the Dirac character of the gauginos (particularly gluinos) can
significantly soften the LHC exclusion limits on direct squark
production \cite{Heikinheimo:2011fk,Kribs:2012gx,Frugiuele:2012kp}.
Moreover, Dirac gluinos above 
the TeV scale are natural \cite{Fox:2002bu}, 
while Majorana gluinos above that scale begin to saturate the
naturalness threshold \cite{Barbieri:1987fn,Papucci:2011wy}. This is a
particularly interesting feature in light of the LHC limits on
gluinos, $ \sim 1.5$ TeV \cite{ATLAS-CONF-2012-109}.

In this paper we will investigate the implications of large flavor mixing 
for the LHC stop phenomenology. 
R-symmetric models provide an optimal setting for this scenario, but our
discussion also applies to the MSSM when a
significant amount of mixing can be accommodated. 
We will focus on a light stop NLSP
(Next-to-Lightest Supersymmetric Particle) 
with a mass splitting $\Delta M = m_{\tilde
t}-m_{LSP}  \lesssim 250$ GeV with the LSP. 
In this region of parameter space  the
mixing induced decays $\tilde{t}\to j+$LSP dominate,
 and hence
completely alter the phenomenology of the stop.

The paper is organized as follows: we briefly review the main features
of models with an approximate R-symmetry. We then present the
effect of large flavor mixing on squark decays, focusing on the stop.  The rest of the paper is devoted to the collider phenomenology of this
scenario, with emphasis on the region of parameter space where the
flavor violating decays dominate.

\section{Mixed stops in R-symmetric models} 

One of the main features of R-symmetric models
is that gauginos are  Dirac fermions instead of Majorana. 
The R-symmetry also forbids $A$ terms, which implies that there is
no left-right mixing. Further, the 
$\mu$-term is also not allowed.
The 
combination of these features relaxes the SUSY
flavor problem, allowing  for a larger flavor violation both in the
squark and in slepton sector than in the MSSM
(see \cite{Kribs:2007ac} and \cite{ Blechman:2008gu}).
For instance, models with Dirac gluinos contribute to meson mixing through dimension 6 operators which are hence suppressed relative to the dimension 5 operators present in models with Majorana gluinos allowing for a potentially large mixing.
 The imaginary
 part of the $K-\bar{K}$ mixing is severely constrained by the measurement
of $\epsilon_K$. Therefore, the mixing in the first and second
generation requires mild flavor structure in order to be consistent
with the limits.

Flavor changing processes like $ \mu \rightarrow e \gamma $
or $b \rightarrow s \gamma$ require a helicity flip in the
corresponding diagrams
which typically would arise from either a Majorana mass insertion,  left
right mixing between the squarks or a $\mu$-term insertion in the large $
\tan{\beta} $ limit; all these terms are forbidden by the R-symmetry.
Therefore, one is left with a mass insertion in the external line
which is significantly smaller, leading to the suppression of all the
$ \Delta F=1 $ processes.
The R-symmetry also protects from dangerous contributions to $ \Delta F=1 $ processes like $ b \rightarrow s \gamma $ arising from a light stop.
Moreover, the dangerous  one-loop contributions to EDMs are forbidden by the
R-symmetry since they all require either a Majorana mass insertion or
squark left-right mixing.

The R-symmetry does not completely solve the flavor puzzle in SUSY. A
mild flavor structure (in particular, a small phase)  is still needed in order
to satisfy the bounds
from $K-\bar{K}$ mixing. However, many flavor violating observables are
absent, and others significantly suppressed. Importantly, these suppressions
are obtained independent of the SUSY breaking/mediation paradigm (beyond the 
requirement of R-symmetry). Therefore, the set of
flavor assumptions for a viable R-symmetric SUSY model is much smaller
than in the MSSM.

 In this
work  we are interested in exploring the consequences of having large flavor mixing on the stop phenomenology.  
As a simplifying assumption we only consider 
  mixing  between  the second or the
first generation and the third generation squarks, both in the left
handed and in the right handed sector.  We assume the unmixed generation to be heavy.
Qualitatively our results continue to hold for more general flavor
structures.

\begin{figure}[tp]
  \begin{center}
    \includegraphics[scale=0.68]{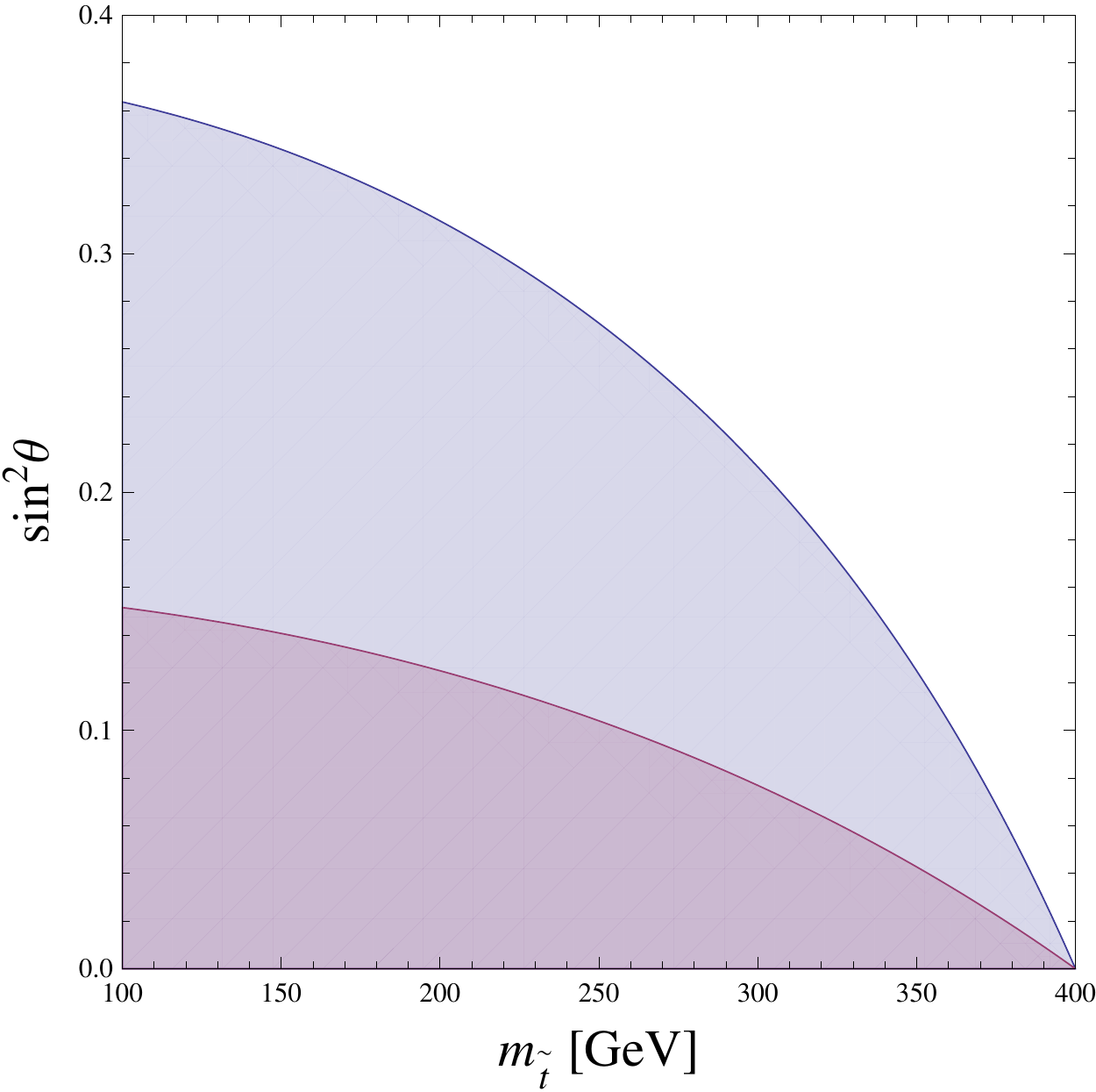}
   
  \end{center}
  \caption{Natural region, which is chosen to be correspond to $(M_{\tilde q})_{33}<$400 GeV, for $m_{\tilde j}=650$ GeV (light purple) and  $m_{\tilde j}=1$ TeV (darker purple).}
  \label{fig:nature}
\end{figure}

The state responsible for canceling the top quadratic divergence is the
``$33$'' state in the squark mass matrix $M^2_{\tilde q}$. Therefore, naturalness
requirements affect $(M^2_{\tilde q})_{33}$, which we choose to be 
$\lesssim (400$GeV)$^2$.
In presence of flavor mixing  $(M^2_{\tilde q})_{33}$ is not 
the physical mass of the (mostly) stop squark, but is related to the 
two mixed squarks masses by the following relation:
\begin{align}
 (M^2_{\tilde q})_{33}
 &= 
 \cos{\theta}^2 m_{\tilde t}^2+ \sin{\theta}^2 m_{\tilde j}^2
\end{align}
where $ \tilde j$ is either a (mostly) charm or up squark.
The Higgs mass parameter thus becomes sensitive to
the potentially heavier squark mass scale through the mixing. Therefore, 
the stop mass is not a robust measure of fine tuning in this scenario, and 
the fine tuning is in general worse than that deduced from the stop mass.
The estimation of fine tuning then involves
both the
mixed generations. This is beyond the scope of this paper, and 
in the following
we will assume the other generations to be above $650$ GeV making them safe from 
the LHC bounds on three degenerate squarks (flavor-mixed partners of $ \tilde t_R$ and $ \tilde t_L$).
This limit can be derived by rescaling the bounds on 8 degenerate squarks with decoupled gluinos.
As shown in figure  \ref{fig:nature}, this spectrum is still
compatible with naturalness ($(M^2_{\tilde q})_{33}< (400$ GeV$)^2$).

It is also important to note that the  presence of a large mixing angle 
can potentially affect the production cross section of stops at the LHC, 
since they can be now produced via their mixing with the charm or up-quark.
However, for sufficiently heavy (Dirac) gluinos we can assume this to be 
a subdominant effect, especially for the case when the stop mixes with 
the charm. Having heavy gluinos also means that the only source of squarks 
production at the LHC (at 7-8 TeV) is direct $ \tilde q^* \tilde q$ production.

\section{Flavor violating stop decay modes}

In this section we discuss the impact of large squark mixing in
the third generation phenomenology, focusing on the
NLSP stop scenario.  This first requires us to specify the nature of the
LSP.  Third generation final states are preferred if the LSP contains
an appreciable Higgsino component. This means that the flavor
violating decays dominate for the first and second generation squarks,
while the third generation phenomenology is not affected by the
mixing.  In this work, we focus on the case 
where the coupling is flavor universal,
and the LSP is a pure bino neutralino or a gravitino.\footnote{We do not consider a wino LSP since it typically implies
light charginos, which are not the focus of our study.}  The gravitino
is often the LSP in R-symmetric models since its mass is related to
the amount of R-breaking communicated to the visible sector through
anomaly mediation \cite{Kribs:2007ac}.  Alternatively, a pure bino LSP
is a good dark matter candidate. It is a pseudo-Dirac fermion, which
means it annihilates more efficiently than the Majorana bino
leading to the correct relic density in a large region of the
parameter space \cite{bettertemp}.
On the other hand at the present day the pseudo-Dirac bino 
behaves like a Majorana
fermion making a it safe from direct detection constraints
\cite{Perelstein:2012qg,Cheung:2012qy}.  
We now discuss the phenomenology of a light stop $ m_{\tilde t} <
400$ GeV for these two scenarios.  
For this purpose it is useful to separately discuss  two
regions: $m_{\tilde{t}}  > m_t + m_{LSP} $  and  $  m_{\tilde{t}}  <
m_t + m_{LSP} $.  

\subsection{$  m_{\tilde{t}}  > m_t + m_{LSP}$}

In this region of the parameter space a stop decays either into a  $
t+$LSP or $ j+$ LSP where the jet is either a charm or up type jet
(for large $ \theta_{23}$ or $ \theta_{13}$ respectively, denoted
$\theta$ in the following).  For a sufficiently heavy stop the
branching ratio for the decay mode into light jet is $
\sin^2{\theta}$, but in the almost degenerate region $m_{\tilde{t}}
\sim m_t + m_{LSP} $ it gets enhanced due to the phase space
suppression of the decay into top quarks, and hence could become the
dominant decay mode.  For instance, for a massless  bino LSP
the branching ratio for these decay modes are:
\begin{align}
  Br(\tilde{t}  \to  j \tilde B) &=  \frac{ \sin^2{\theta} }{ \sin^2{\theta}+  \cos^2{\theta}  (1-\frac{m_t^2}{m_{\tilde t}^2 })^2  } , \\
  Br(\tilde{t}  \to  t  \tilde B) &=  \frac{ \cos^2{\theta}  (1-\frac{m_t^2}{m_{\tilde t}^2})^2}{ \sin^2{\theta}+  \cos^2{\theta}  (1-\frac{m_t^2}{m_{\tilde t}^2 })^2  }, \
\end{align}
and for a massless gravitino LSP:
\begin{align}
  Br(\tilde{t}  \to  j G) &=  \frac{ \sin^2{\theta} }{ \sin^2{\theta}+  \cos^2{\theta}  (1-\frac{m_t^2}{m_{\tilde t}^2 })^4  } , \\
  Br(\tilde{t}  \to  t  G) &=  \frac{ \cos^2{\theta}  (1-\frac{m_t^2}{m_{\tilde t}^2})^4}{ \sin^2{\theta}+  \cos^2{\theta}  (1-\frac{m_t^2}{m_{\tilde t}^4 })^4  }.\
\end{align}
These equations apply for decays of both $ \tilde t_L$ and $ \tilde t_R$.
\begin{figure}[tp]
  \begin{center}
    \includegraphics[scale=0.9]{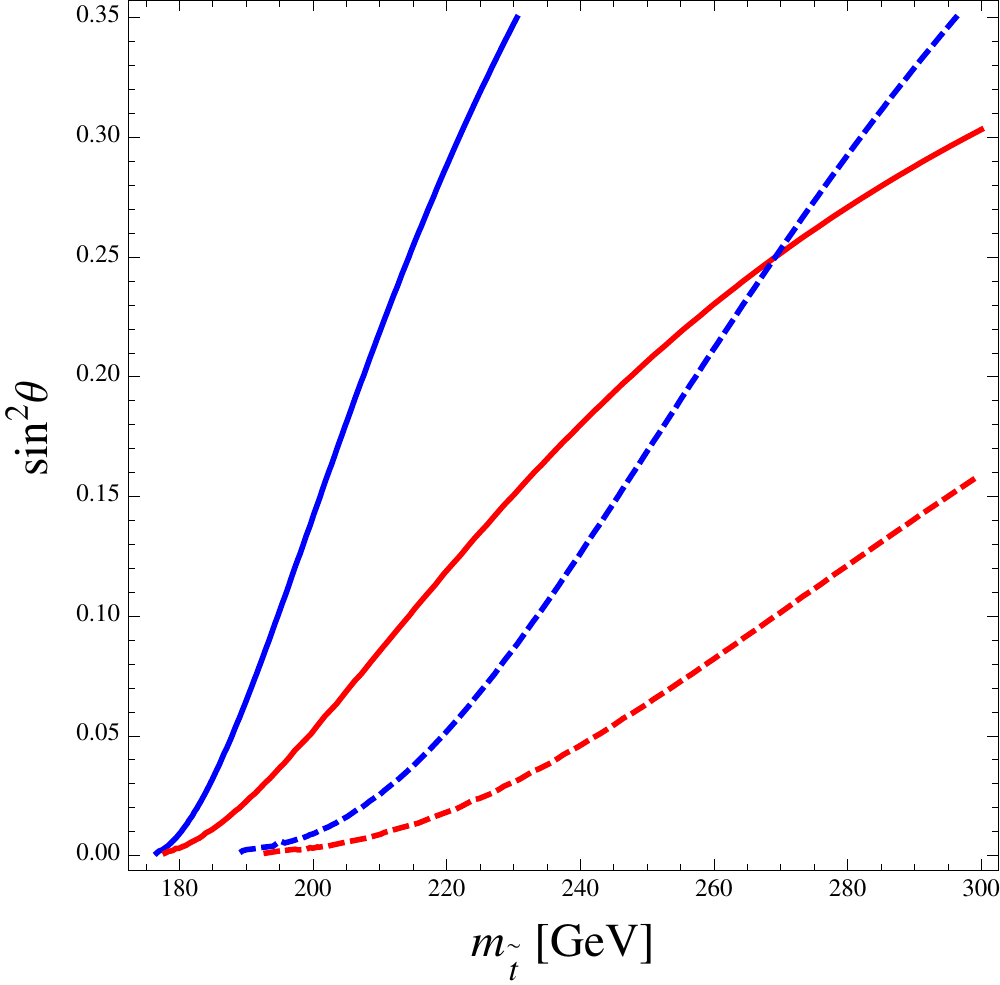}
     \end{center}
  \caption{Contour plot for the branching ratio $ Br(\tilde t \to \tilde B+j$)  (solid) and  Br($ \tilde t \to G+j$) (dashed) as function of the stop mass and the mixing. Contours are shown for branching ratio 75 \% (blue)  and 50\% (red). The LSP is taken to be massless in both cases. }
  \label{fig:BRgr}
\end{figure}

As shown in  figure \ref{fig:BRgr} 
the FV decay mode dominates over $t+$LSP for mass splitting $ \Delta M \sim m_t$
 due to the large
mixing angle and phase space suppression. 
For a mixing of $ \sin^2{\theta}=0.25$ the FV has order one branching ratio for stop masses up to 250 GeV for  a massless bino LSP.
For the gravitino this mode continues to dominate for 
much smaller mixing angles.
Therefore, the FV mode is significant up to masses where dedicated searches for the topology $t+$LSP starts to be sensitive.

This is interesting since the region $m_t< m_{\tilde t}<$ 250 GeV
represents an important lack of coverage of  LHC  analyses. Indeed the sensitivity of all the stop dedicated searches
in the channel $ t $ plus LSP starts at 230-250 GeV both for ATLAS
\cite{ATLAS-CONF-2013-024} and CMS \cite{CMS-PAS-SUS-12-023} for a
massless LSP. 
The challenge here is in distinguishing a possible stop signal over the top
background.
Due to the experimental difficulties this region 
($m_{\tilde t} \sim m_{t}$)  is called the stealth stop region
\cite{Han:2012fw}. For proposed 
dedicated searches in this region see \cite{Alves:2012ft,Kilic:2012kw}.
Alternatively, one can hope to discover the stealth stop relying on a
different decay mode.
In the MSSM  when the stop is the NLSP the only other possible decay
mode is into an off shell top and the
LSP. However, the branching ratio for this three-body decay ends
abruptly at $ m_{\tilde t}=m_t$ for a neutralino LSP. If the gravitino
is the LSP
instead the branching ratio for three-body decay is significant ($\sim 20 \%$) up to 200 GeV, so the stealthy region is less broad \cite{Han:2012fw}. 
We have shown  that in our scenario, if the mixing is significant,  a
stop almost degenerate with the top quark will decay mostly into a
light jet plus LSP.  This can simplify the challenge of looking for
stops into the stealth region.  We will discuss this in the next
section. 
 
For heavier stops, $m_{\tilde t} \gg m_{t}+m_{LSP},$  the branching
ratio for the decay mode  into light jet is still large, but
subdominant (it is proportional to $ \sin^2{\theta}$).  As
discussed in  \cite{Blanke:2013uia}  this can potentially relax the
bounds on the stops from the $ t$ LSP searches,
\cite{ATLAS-CONF-2013-024} and \cite{CMS-PAS-SUS-12-023}.
 Another interesting phenomenological implication of a significant squark mixing is single top production at the
LHC ($ p p \to \tilde t^* \tilde t$ with $ \tilde t \to j+$LSP and $
\tilde t^* \to t^*+$ LSP and vice versa).  This topology constitutes a
``smoking gun'' signature for scenarios with a large squark mixing and
it has been discussed in   \cite{ Blanke:2013uia}  and  more
extensively in \cite{Kribs:2009zy}.  It is important to note that the
CMS and ATLAS analysis  do not have a dedicated analysis for this
mixed decay topology despite the low SM background.

 \subsection{$ m_{\tilde{t}}  < m_t + m_{LSP} $}
 The two-body
flavor conserving stop decays are forbidden for  $  m_{\tilde{t}}  <
m_t + m_{LSP} $.  We will show that, for sufficiently large mixing
angle, the two-body flavor violating mode dominates over the
three-body decay $ \tilde t \rightarrow b\ W$ LSP.

\begin{figure}[tp]
  \begin{center}
    \includegraphics[scale=0.74]{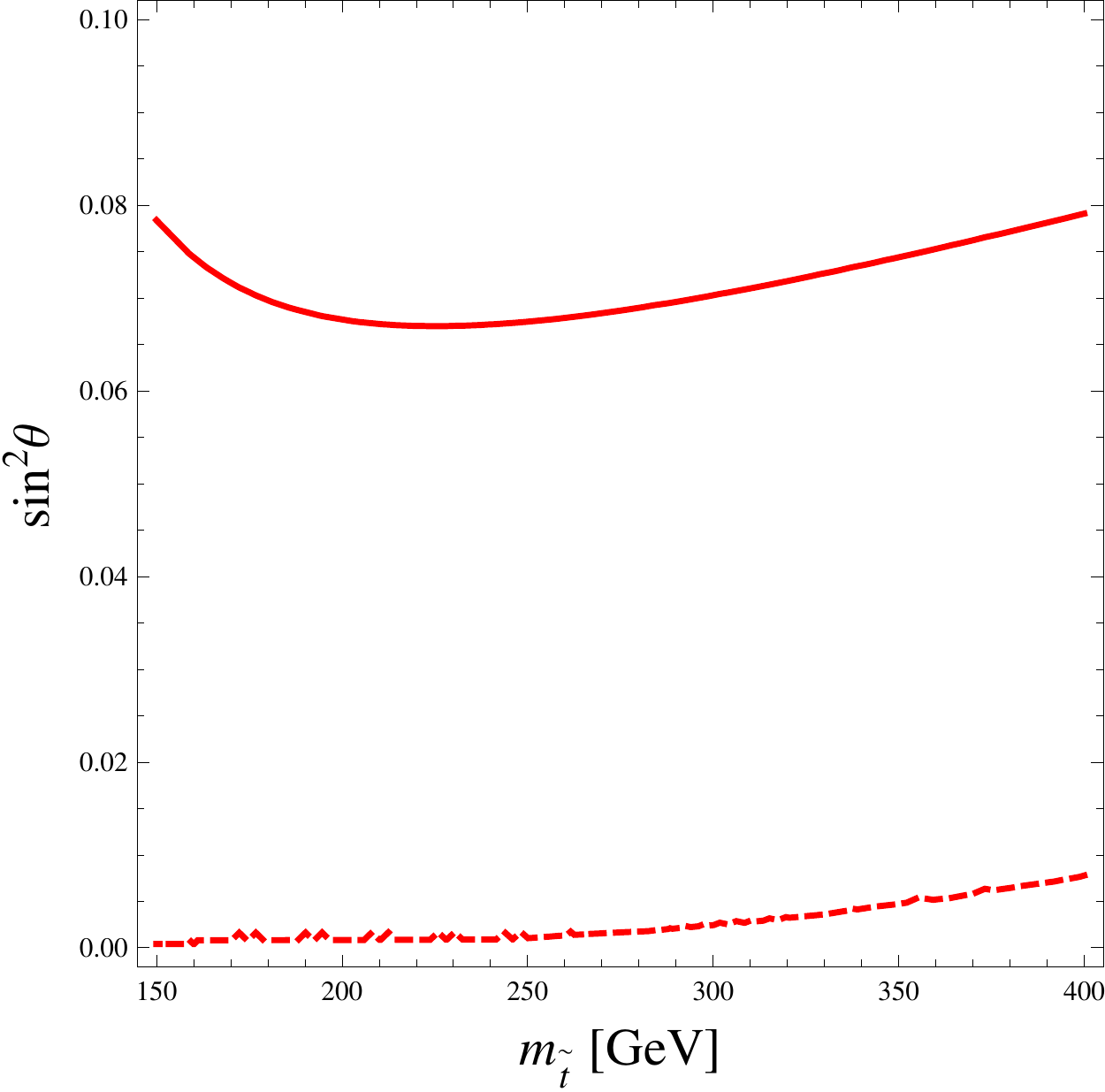}
  \end{center}
  \caption{Contour plot for $ 90 \%$ branching ratio for the two-body
  decay $\tilde{t}_L \to j LSP$ for a bino LSP (solid) and a
  gravitino LSP (dashed), for $m_{\tilde t} - m_{LSP}=150$ GeV. 
  A similar plot applies to the $\tilde t_R$
  decay modes.}
  \label{fig:BRstoplight}
\end{figure}

If the gravitino is the LSP then the decay width for  two open decay modes
are \cite{Kats:2011it}: 
\begin{align}
  \Gamma(\tilde t \rightarrow b W G) 
  & \sim 
  \cos^2{\theta} 
  \frac{\alpha }{ \sin^2{\theta_W}} 
  \frac{ (m_{\tilde t}-m_W)^7}{128 \pi^2 m^2_W F^2} 
  ,\\
  \Gamma(\tilde t \rightarrow j G) 
  & \sim
  \sin^2{\theta}
  \frac{m_{\tilde t}^5 }{16 \pi F^2}  ,
\end{align}
where $\sqrt{F}$ is the SUSY breaking scale. The branching ratio
for the flavor violating (FV) two-body decay is:
\begin{align}
  Br(\tilde t \to j  G) =  \frac{1}{1+ \cot{\theta}^2  (\frac{ \alpha m^2_{\tilde t} (1-\frac{m_w}{m_{\tilde t}})^7}{8 \pi \sin^2{\theta_W} m_W^2 })}
\end{align}

When the gravitino is the LSP the two-body decay mode dominates over
three-body decays even for very small mixing angles, as 
shown in figure \ref{fig:BRstoplight}.
This decay mode is prompt for a low SUSY breaking scale 
($\sqrt{F} \lesssim 100$ TeV).
For these SUSY breaking scales the gravitino is extremely light and
hence in the following we will take it to be massless.
If the LSP is purely bino, the
two-body decay dominates (with branching ratio greater than $90 \%$)
over the three-body decay for a larger mixing angle
(around $ \sin^2{\theta} \sim 0.08$) than in the gravitino case 
as it shown in
figure \ref{fig:BRstoplight} for  a left handed stop (the same conclusion
applies to $\tilde t_R$). We conclude then
that the FV two-body decay mode is the dominant decay mode for the
stop in the entire region of the parameter space where $m_{\tilde{t}}
< m_t + m_{LSP}$ for moderate mixing.  Also, as shown
above, the region where it dominates extends above the 
$m_{\tilde{t}} =  m_t + m_{LSP}$ threshold into the so-called stealth
region.  Therefore, the decay mode into jet plus LSP is a potential discovery
mode for a stop NLSP almost
degenerate with the LSP.  
It is interesting to
briefly compare the FV stop decay mode in our scenario with the MSSM
with minimal
flavor violation (MFV).  In the MSSM the dominant decay mode is the
3-body
decay into $ b W $ plus the LSP, and when this is kinematically closed
($m_{\tilde{t}} -m_{LSP}<m_W+m_b$) the dominant decay mode is
either the loop suppressed two-body flavor violating (FV) decay mode,
$ \tilde t \to c+ $LSP or the four body decay.  The FV stop decay mode
is suppressed compared to the three-body decay  since the effective
coupling $ \tilde t \  \chi_0^1 \ c $ is of order $ 10^{-5}$ assuming
MFV (see \cite{Han:2003qe,Carena:2008mj,Muhlleitner:2011ww}).

\section{LHC sensitivity to a light stop NLSP  with large flavor
mixing} 

In the previous  section we showed that the FV decay mode
into a light jet (either a charm or up-quark initiated) and the LSP is
the dominant
decay mode and hence a potential discovery mode for a stop almost
degenerate with the LSP when a moderate squark flavor mixing is
present.  It is therefore interesting to determine the LHC
sensitivity for this topology, and the bounds on the stop
mass in this region of the parameter space.

The Tevatron already puts interesting constraints on stops
decaying into charm jets.
Both D0 \cite{Abazov:2007ak} and CDF \cite{Aaltonen:2012tq}
have performed a
dedicated analysis for $ \tilde t \rightarrow c+$LSP,  putting a
bound on the stop mass as strong as 180 GeV for a 90 GeV LSP.
However, both the more compressed region $\Delta M<40$ GeV  and the
region outside $m_{\tilde{t}}-m_{\tilde B}<m_W+m_b$, were left
unconstrained.  These analyses rely on $c$-tagging, and hence
do not directly constrain the FV stop decay into an up quark.

ATLAS and CMS do not have any dedicated analysis for this topology ---
light stops are searched in the $\tilde \chi_1^{\pm} b$ channel
\cite{ATLAS-CONF-2012-167}.  Nonetheless, since our final state
contains missing energy and two jets, jets plus missing energy (MET)
LHC searches are potentially sensitive to this topology.  Limits on
our case can be derived from simplified models for squark searches
with decoupled gluinos.  However, CMS and ATLAS analyses do not extend
to the region of the parameter space significant for us, that is $
m_{\tilde t} <300$ GeV and/or $\Delta M= m_{\tilde t}-m_{LSP}< m_t$. The
reason is that signal efficiency relies in part on the presence of
initial state radiation which is associated with large uncertainty.
Therefore, our model is largely unconstrained by the LHC analyses,
allowing for a stop as light as $100$ GeV, the bound from LEP
\cite{Abbiendi:2002mp}. Dedicated searches
for $\tilde{b}_L \to b+$LSP can potentially 
constrain the left-handed stop significantly. In this paper we will
focus on the LHC sensitivity to the flavor mixed stop signals only.

Beyond this particular topology, in general there is an insufficient
coverage of the
phenomenology of a light stop somewhat degenerate with the LSP in a
large region of the available parameter space.
Motivated by this lack of coverage several external analyses 
\cite{Krizka:2012ah,Delgado:2012eu,Dreiner:2012sh,Dreiner:2012gx,Yu:2012kj}
have shown that LHC is sensitive to a broader region of the parameter 
space than the one explored by the public analyses.
\cite{Krizka:2012ah} and \cite{Delgado:2012eu} analyze the LHC
sensitivity for the stop four body decay, and a bound on its mass  of
around 250 GeV was estimated. 
Furthermore, \cite{Krizka:2012ah} discuss the FV stop decay,
focusing on the region  $m_{\tilde{t}}-m_{\tilde B}<m_W+m_b$ where the
FV mode could dominate even in the MSSM.
In \cite{Dreiner:2012sh} a model independent study was performed, and the FV stop decay into a light jet plus
the LSP was investigated beyond this particular compressed region.
A bound of  $\sim 300$ GeV on the stop mass was derived from
shape-based hadronic analyses  and monojet searches
\cite{ATLAS-CONF-2011-096,Chatrchyan:2012me}.

In this section we will study a simplified model where the stop
decays with a 100\% branching ratio to $j+$LSP. 
We simulate the stop production in MadGraph5 \cite{Alwall:2011uj}. The stops are then
decayed and the decay products hadronized in Pythia \cite{Sjostrand:2006za}, which also adds
initial and final state showers. Since we are interested in hard jets
arising from this radiation, we simulate $\tilde{t}^* \tilde{t}$,
$\tilde{t}^* \tilde{t}+j$ and $\tilde{t}^* \tilde{t}+2j$ at the matrix
element level and match it with the Pythia shower. We use PGS to
simulate detector effects on our signal. The detector simulation and the QCD modeling  are
probably the strongest limiting factors of our simulation. We emulate relevant LHC analyses for
our topology.

\subsection{Jets+MET searches} 

Traditional jets+MET searches both of ATLAS and CMS are designed for
QCD production for new
physics particles such as squarks or gluinos, which subsequently decay
into jets and the LSP -- which escapes detection -- leading to a
missing energy signature. 

For spectra with a mass large splitting with the LSP, these searches
are very effective.  The limits on squark simplified models are
crossing the TeV threshold.  However, these limits
are considerably weakened if only one quark eigenstate is light
\cite{Mahbubani:2012qq}, even if the LSP is massless.  The sensitivity of
these searches is much worse for the compressed region (for instance,
for our region of interest,
$m_{\tilde{q}} - m_{LSP} \lesssim m_t$).  Most of the
data for these searches comes from high instantaneous luminosity
samples. Consequently, the hadronic triggers relevant for these
searches require a very high amount of hadronic activity ($H_T + E_T
\gtrsim 1200$ GeV for relevant signal regions in the $4.7$ fb$^{-1}$
ATLAS analysis, and $ H_T > 500$ GeV for the $4.98$ fb$^{-1}$ CMS sample
\cite{:2012mfa}). Therefore, these searches are very insensitive to
our region of interest.  The dedicated searches for the sbottom
decaying into $ b+$LSP \cite{ATLAS-CONF-2012-165} could be potentially sensitive
to our topology since a charm has a $ 10 \% $
probability to be mistagged as a $b$-quark.  It would be then be
interesting to recast the dedicated searches for third generation
squarks as done in \cite{Krizka:2012ah} , but extending the region of
the parameter space.  The possibility of a large squark mixing with a
$c$-quark final state could provide a motivation for an improved $c$-tagging.

\subsection{Shape-based searches} 

Shape-based analyses like the CMS $\alpha_T$ and the CMS razor
analyses \cite{Chatrchyan:2012uea,CMS:2012dwa,CMS:2012cwa} are much
more sensitive to our topology.
Instead of only using the overall energy scale in the event, these
searches also employ an event shape-based discriminant. This allows
the use of shape-based triggers requiring much lower $H_T$ thresholds.
We derive limits in the stop-LSP mass plane for a large region of
parameter space.  We scan over the entire plane (beyond the compressed
region) assuming a 100\% branching ratio to jet + LSP. 
While this might be an oversimplification for a realistic model
(for example, the single top
channel, if open, might provide stronger limits), it allows us to
investigate the efficiency of this particular topology in a model
independent way.

\subsubsection{$\alpha_T$ analysis} 

The CMS SUSY hadronic searches employing $\alpha_T$
\cite{Chatrchyan:2012wa} are sensitive to events with 2 or more
energetic jets with missing energy. The
$\alpha_T$ variable was first discussed in \cite{Randall:2008rw}.
Since the additional trigger on $\alpha_T$ allows the $H_T$  and $p_T$
thresholds to
be significantly lowered,  these searches are more
sensitive to our region over traditional jets+MET searches. 

In this type of analysis all events are clustered into a dijet
topology containing two pseudo-jets, choosing the combination which
minimizes the $E_T$
difference between the two pseudo-jets.

The variable $\alpha_T$ is defined as,
\begin{align}
  \alpha_T
  &=
  \frac{E_T^{j_2}}{M_T}
  \qquad\qquad
  &M_T
  &=
  \sqrt{
  (E_T^{j_1}+E_T^{j_2})^2
  -(p_x^{j_1}+p_x^{j_2})^2
  -(p_y^{j_1}+p_y^{j_2})^2
  }
\end{align}
where $E_T$ for each pseudo-jet is defined as the scalar sum of $p_T$
of its components.

Jets considered in the analysis are
required to have $E_T>50$ GeV.  Further, the two highest-$E_T$ jets
must satisfy $E_T>100$ GeV each, and the highest $E_T$ jet should be
central, $|\eta|<2.5$. Events with additional hard jets ($E_T>50$ GeV)
are vetoed if the jet $|\eta|>3.0$. Events are also required to have
significant hadronic activity $H_T > 275$ GeV. Finally, a requirement
of $\alpha_T>0.55$ brings the QCD backgrounds down to manageable
levels.  Events which pass these requirements are categorised
according to their $H_T$. The $p_T$ thresholds in the lowest two $H_T$
bins are rescaled to keep the efficiency of the $\alpha_T$ cut similar
in each $H_T$ region.

\begin{figure}[tp]
  \begin{center}
    \psfragfig*[width=0.7\textwidth]{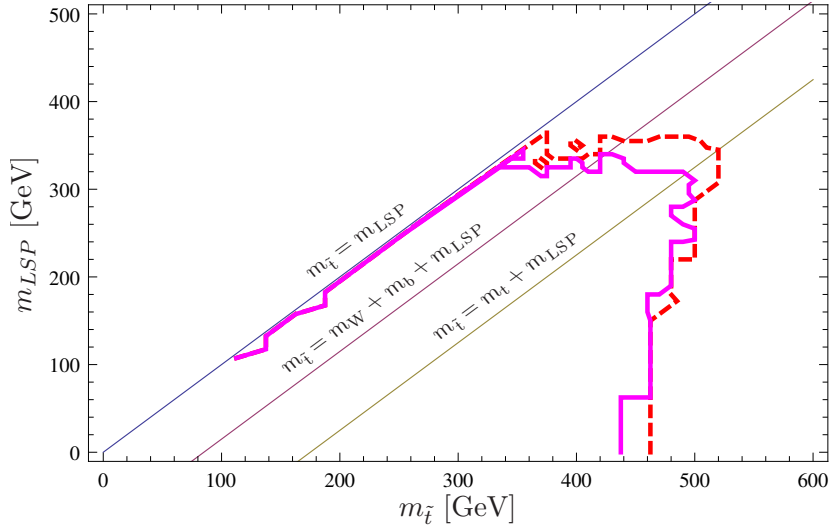}{
    \psfrag{M}[c][t]{$m_{\tilde{t}}$ [GeV]}
    \psfrag{N}[c][c]{$m_{LSP}$ [GeV]}
    \psfrag{A}[r][c][0.8]{$m_{\tilde t} = m_{LSP}$}
    \psfrag{B}[r][c][0.8]{$m_{\tilde t} = m_W + m_b + m_{LSP}$}
    \psfrag{C}[r][c][0.8]{$m_{\tilde t} = m_t + m_{LSP}$}
    }
  \end{center}
  \caption{Limits from the 7 TeV  $\alpha_T$ analysis in the $(m_{\tilde t},m_{LSP})$ plane.
  Dashed red lines indicate observed limit, while the solid magenta
  line shows the expected limit.
  The stop is assumed to decay into jet and LSP with $100\%$  branching ratio everywhere.}
  \label{fig:alphaT}
\end{figure}

\begin{figure}[tp]
  \begin{center}
    \psfragfig*[width=0.7\textwidth]{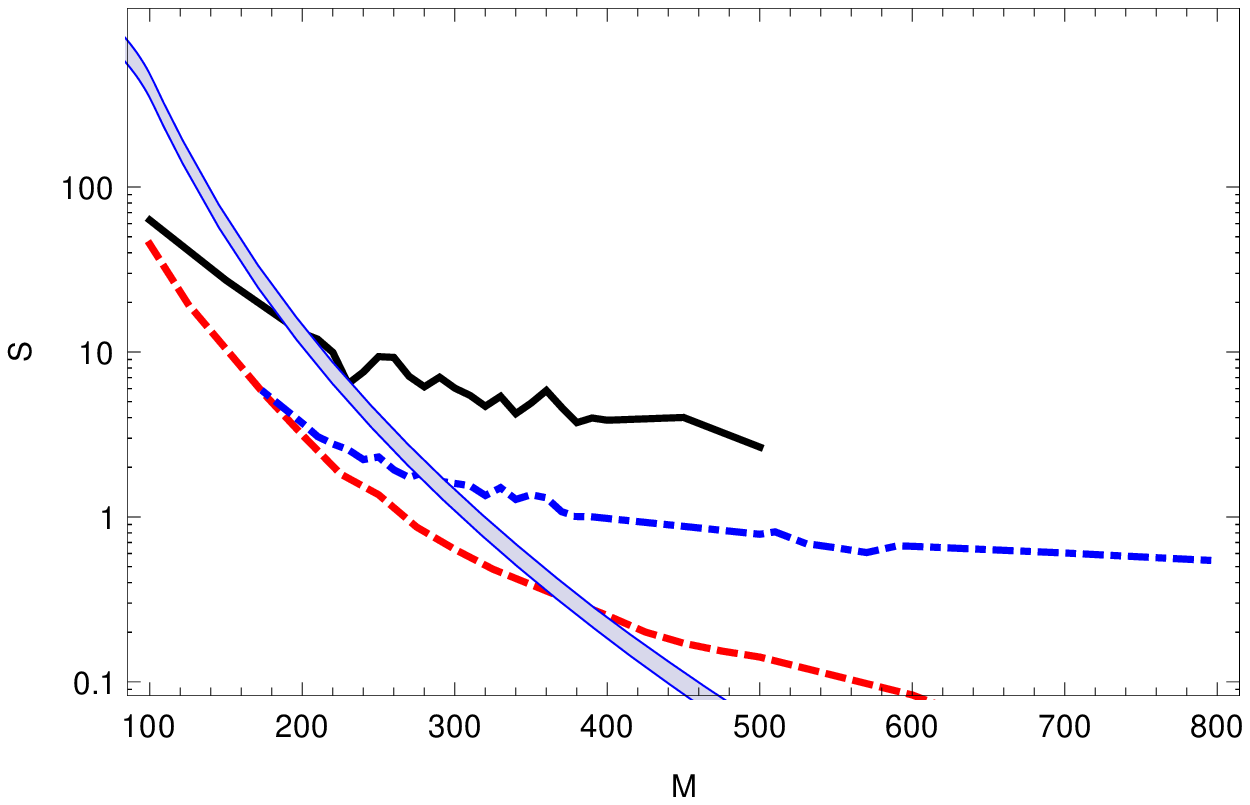}{
    \psfrag{M}[c][c]{$m_{\tilde{t}}$ [GeV]}
    \psfrag{S}[c][c]{$\sigma_{\tilde{t}^*\tilde{t}}$ [pb]}
    }
    \vspace{15mm}
    \\
    \psfragfig*[width=0.7\textwidth]{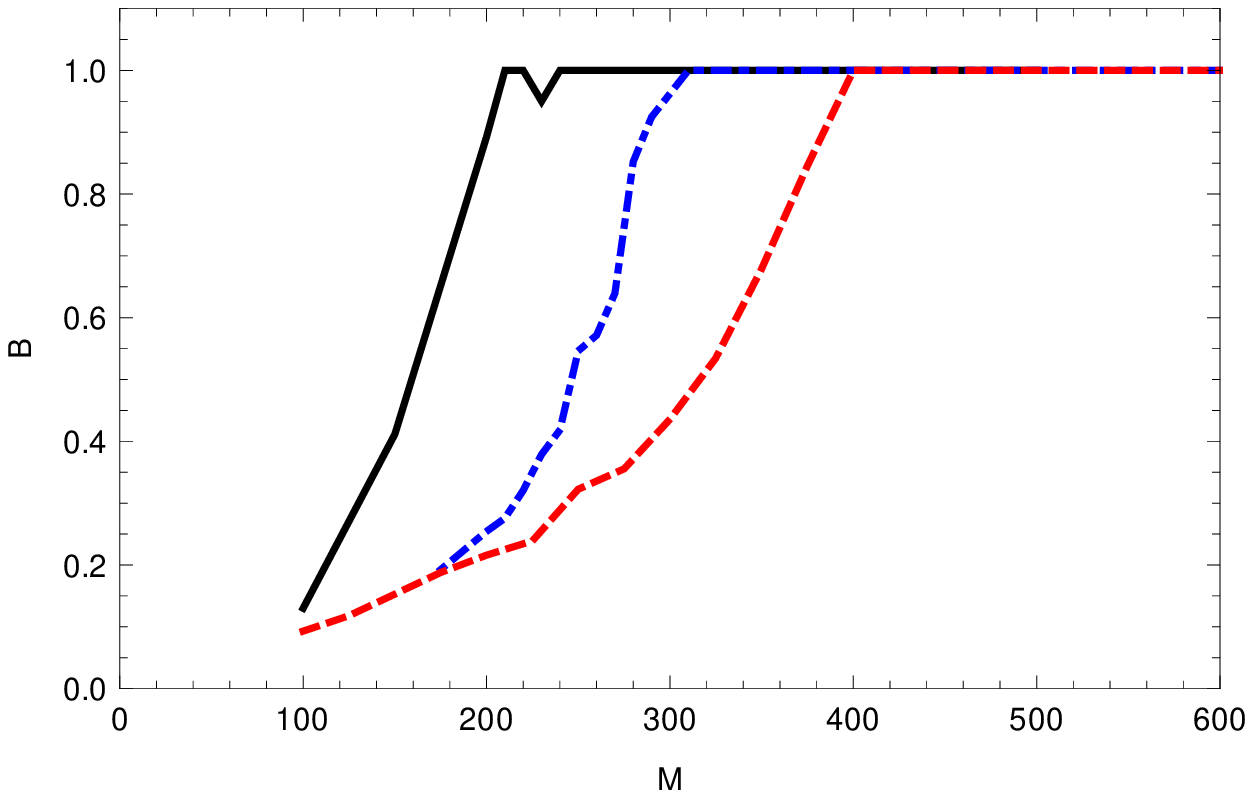}{
    \psfrag{M}[c][c]{$m_{\tilde{t}}$ [GeV]}
    \psfrag{B}[c][c]{Br$(\tilde{t}\to j+$LSP)}
    }
  \end{center}
  \caption{
  Cross section (top) and branching ratio (bottom)
  limits from the 7 TeV $ \alpha_T$ analysis as a function of stop mass for various LSP masses,
  (solid black $m_{\tilde{t}}-m_{\tilde{B}}=10$ GeV,
  dot-dashed blue: $m_{\tilde{t}}-m_{\tilde{B}}=150$ GeV,
  dashed red: $m_{\tilde{B}}=0$ GeV and $m_{3/2}=0$ GeV). 
  The gray band is the NLO stop production
  cross section at 7 TeV  computed using Prospino 2.1 \cite{Beenakker:1996ed}.}
  \label{fig:alphaT1}
\end{figure}

The CMS analysis   \cite{Chatrchyan:2012wa}
does not cover the parameter space we are
interested in. We emulate the $\alpha_T$ analysis on our
simulated sample (the limits are presented in figure 
\ref{fig:alphaT} and figure 
\ref{fig:alphaT1}). 
The experimental limit is calculated using a likelihood model to test
for the presence of a variety of signal models. Here we will estimate
the $95\%$  upper limit on the cross section by the following formula,
\begin{align}
  \tilde \chi^2 
  &=
  \frac{(\Delta N - L \sigma \epsilon)^2}
  {N_{SM} + \delta_{SM}^2 + L \sigma\epsilon}
  = 3.84
\end{align}
where $\epsilon$ is the signal efficiency, $L=4.98$ fb$^{-1}$ is the luminosity,
$N_{SM}$ is the expected number of events from SM and $\delta_{SM}$ is
the uncertainty in the expected number. The only relevant channel for this search is
the 0 b-quark jets channel. It turns out that a few bins have downward 
fluctuations in the background, and hence 
tend to lead to a tighter limit. To be conservative, we
present both the observed and expected limits. We define $\Delta N$ as,
\begin{align}
  \Delta N
  &=
  \left\{
  \begin{array}{cc}
    0 &\text{expected limit}\\
    N_{obs}-N_{SM} & \text{observed limit}
  \end{array}
  \right.
\end{align}
Further, we do not attempt to combine limits from different bins, but
consider a parameter point ruled out at 95\% confidence if the
production cross section for that point exceeds the limit from
\emph{any} individual bin.

Our  $95\%$  level  limits, computed for a single stop eigenstate
decaying a 100\% into jet and the LSP, are presented in the stop-LSP mass
plane in figure  \ref{fig:alphaT}. Our estimate excludes a single stop
$ \tilde t $  below 250 GeV in the highly compressed region $ \Delta M
< m_W+m_b+ m_{LSP} $, while for an increased mass splitting ($ \Delta
M \sim 150$ GeV) the limit is around 350 GeV. Stop masses below $
\sim 420$ GeV  are excluded for massless LSP. The interpretation of
the actual limit within a realistic model with mixing is not
straightforward since for
$ \Delta M > m_t$ the decay mode into $j+$LSP is no longer the dominant
decay mode, and therefore the branching
ratio suppression has to be taken into account.

The left plot in figure  \ref{fig:alphaT1} shows the $ 95 \%$ limit on
the cross section for different mass splittings, $ 200$ GeV  $<\Delta
M <$ 250 GeV.  Also, we interpret these as  limit on the branching
ratio as a function of the stop mass. We notice that the cross section
limits are relatively flat in $m_{\tilde{t}}$, while the stop
production cross section is steeply falling. Thus, the branching ratio
suppression does not result in a significantly different limit.
At the same time, when we add additional states at the same mass (for
instance, $ \tilde t_R, \tilde t_L $ and $\tilde b_L$ all together or
just $\tilde t_L $ and $ \tilde b_L$) the cross section gets
multiplied by a factor of 2 or 3. By the same logic, these limits do
not get much worse in this case. 

\begin{figure}[tp]
  \begin{center}
    \includegraphics{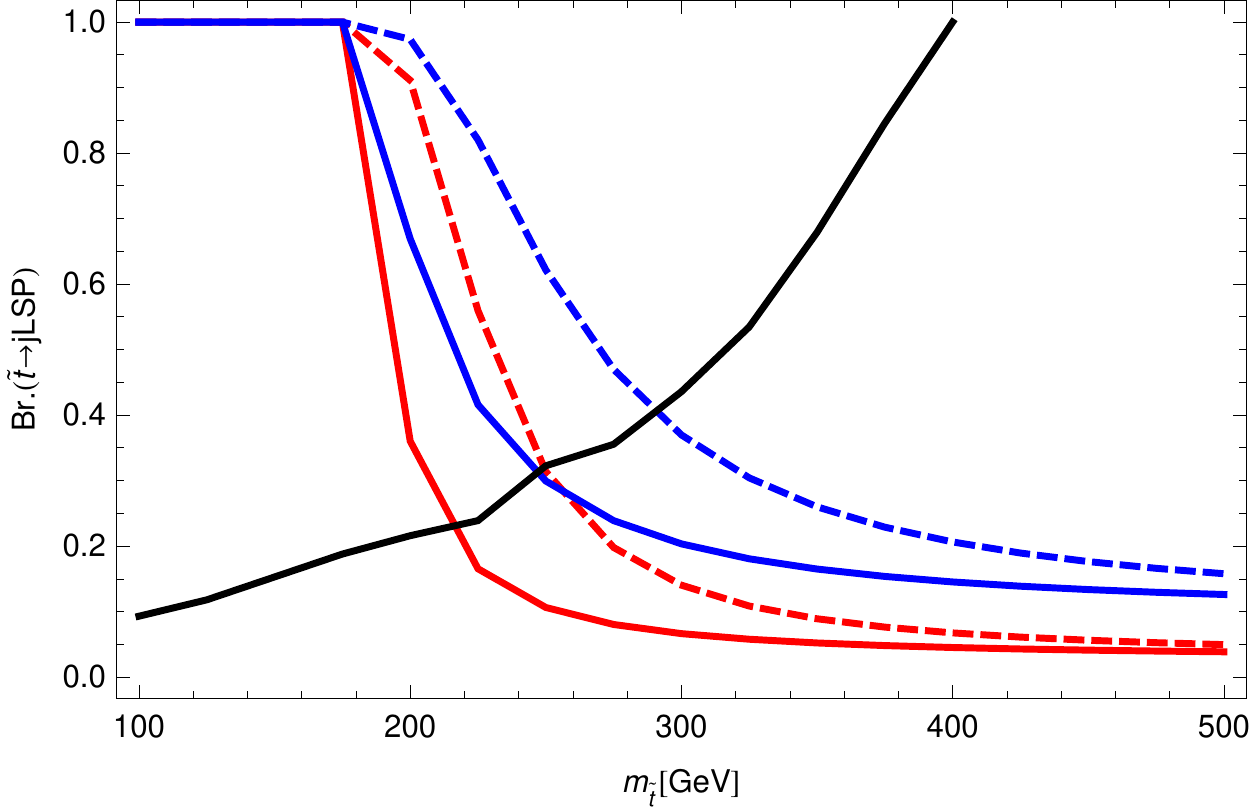}
  \end{center}
  \caption{The limit on the branching ratio $\tilde{t}\to j LSP$ as a function of the stop mass (black solid).
    The different curves represent branching ratios for  $\tilde{t}\to j \tilde G$ (dashed) and $\tilde{t}\to j \tilde B$ (solid) considering  $\sin^2{\theta}=0.03$ (red) and 0.1(blue).
   The LSP is always assumed to be massless.}
  \label{fig:gravitino}
\end{figure}

The plot in figure  \ref{fig:gravitino}  shows the limits on the stop
mass  in the region $ \Delta M > m_t$ for a massless LSP for different
values of the mixing angle. For a gravitino LSP, for mixing as low as
$3 \%$ the entire stealth region $ m_{\tilde t} \lesssim 250 $ GeV is
excluded.  For a massless bino instead, the stealth region  is ruled
out for a larger mixing angle $ \sin^2{\theta} \sim 0.1$.  
For a heavier stop, dedicated searches for $t+$LSP decay mode
  start to be sensitive setting  a limit which depends on the amount
  of mixing.  Stops of mass up to $500$ GeV are excluded even in the
  maximal mixed scenario  \cite{Blanke:2013uia}. Therefore, the
  natural window for a massless LSP and a mixed stop could be almost
  entirely covered.

\subsubsection{Razor analysis} 

The razor is an inclusive analysis which divides the signal
and background into various boxes based on the final states. A global
limit can then be derived from a likelihood analysis over every box.
The power of the analysis lies in the fact that the backgrounds in
each of the boxes are well-fit by a simple function, hence allowing for a
relatively small error in the background extrapolation into the signal
region.  In our case, the signal only populates the HadBox (defined
below), and hence all limits will be derived from this box only.

The razor search is designed to look for pair production of new heavy
states which yield high energy jets. All events are clustered into two
megajets to reduce every final state into the dijet topology. The two
razor variables employed, $M_R$ and $R$ are then calculated from these
megajets. The $M_R$ razor kinematic variable is given as,
\begin{align}
  M_R
  &=
  \left[
  (|\vec{p}^{j_1}|+
  |\vec{p}^{j_2}|)^2
  -(p_z^{j_1} + p_z^{j_2})^2
  \right]^{\frac12}
\end{align}
and is invariant under longitudinal boosts.
$M_R$ approximates the energy scale in the
heavy particle decay and peaks at,
\begin{align}
  M_\Delta
  &=
  \frac{m_{\tilde{t}}^2 - m_{LSP}^2}{m_{\tilde{t}}}
\end{align}

The $R$ variable is defined as 
\begin{align}
  R= M_T^R / M_R ,
\end{align}
where $M_T^R\leq
M_\Delta$ is a transverse invariant mass, defined as,
\begin{align}
  M_T^R
  &=
  \left[
  \frac12 \left(
  E_T^{miss}(p_T^{j_1} + p_T^{j_2}) 
  - \vec{E}_T^{miss}\cdot(\vec{p}_T^{j_1} + \vec{p}_T^{j_2})
  \right)
  \right]^{\frac12}
\end{align}

QCD jets typically have a small value of $R$, and the QCD distribution
peaks at $R\sim0$, whereas the heavy particle production can
produce higher values of $R$.

In the compressed region, much of the jet activity arises from QCD
radiation, and hence the above observations are not directly
applicable. The $M_R$ and $R^2$ distributions depend weakly on the
stop mass, but are largely determined by the jet $p_T$ selection. 
Therefore, it is remarkable that the razor searches are effective in
this region of parameter space. In fact, we find that the razor analysis limits
go all the way to the degeneracy limit ($m_{\tilde{t}} = m_{LSP}$), where
the stop decay products are not visible, and all hadronic activity
arises purely from initial and final state radiation. It has been
pointed out that in this region, the stop search is analogous to the
dark matter search (\cite{Delgado:2012eu,Fox:2012ee}). For the same
reason, the monojet analysis designed for dark matter search
sets significant limits on the stop masses for a very
compressed spectrum ($\Delta M<30-40$ GeV)
\cite{Krizka:2012ah,Delgado:2012eu,Dreiner:2012sh,Dreiner:2012gx}.

As mentioned above, the only box relevant for our simplified
model is the HadBox. An event is put into the HadBox if it does not
fall into any other Box (involving electron and/or muon final
states), and satisfies $M_R>400$ GeV and $0.18 < R^2 < 0.5$,  \cite{Chatrchyan:2012uea}.

The crucial feature of this search driving the limits in our case is
the relatively low $p_T$ thresholds. All jets with $p_T>40$ GeV and
$|\eta|<3.0$ are considered. There is a further requirement on the two
highest-$p_T$ jets to have $p_T>60$ GeV.

The limits are derived according to the procedure outlined in  \cite{Delgado:2012eu}. From each box (which in our case is just one box), a posterior-probability density function is derived for the signal cross section,
\begin{align}
  P(\sigma)
  &=
  \int_0^\infty db \int_0^1 d\epsilon
  \frac{(b+L\sigma \epsilon)^n e^{-b-L\sigma\epsilon}}{n!}
  \text{lognormal}(\epsilon|\bar{\epsilon},\delta_{\epsilon})
  \text{lognormal}(b|\bar{b},\delta_{b})
\end{align}
where $b$ is the actual background yield, $\bar{b}$ is the expected
background yield and $\delta_b$ is the error in the background modeled
using a lognormal distribution. The corresponding quantities for
$\epsilon$ denote the efficiencies. An error of $30\%$ was chosen for
the signal efficiency to account for detector simulation errors.
$\sigma$ is the cross section, $n$ the observed number of events and
$L$ is the available luminosity. The $95\%$ limit is then
straightforwardly derived from the cumulative distribution of
$P(\sigma)$.
\begin{figure}[tp]
  \begin{center}
    \psfragfig*[width=0.7\textwidth]{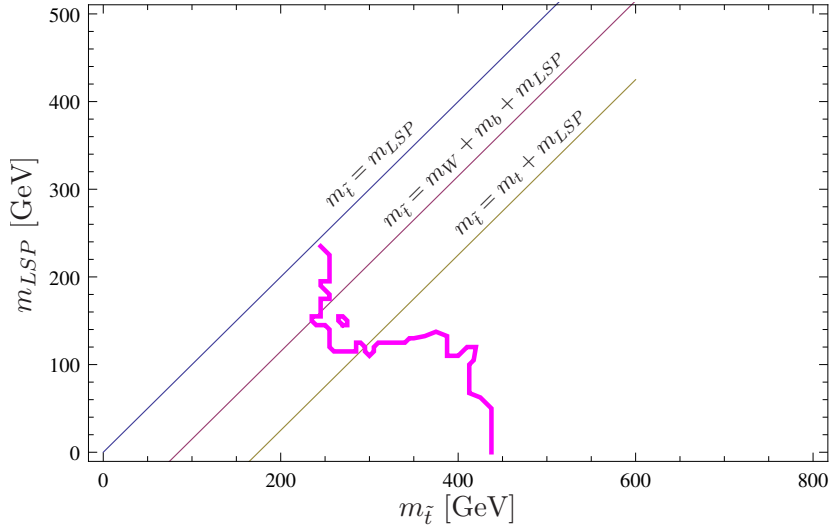}{
    \psfrag{M}[c][t]{$m_{\tilde{t}}$ [GeV]}
    \psfrag{N}[c][c]{$m_{LSP}$ [GeV]}
    \psfrag{A}[l][c][0.8]{$m_{\tilde t} = m_{LSP}$}
    \psfrag{B}[l][c][0.8]{$m_{\tilde t} = m_W + m_b + m_{LSP}$}
    \psfrag{C}[l][c][0.8]{$m_{\tilde t} = m_t + m_{LSP}$}
    }
  \end{center}
  \caption{Limits on our scenario from the 7 TeV razor analysis \cite{Chatrchyan:2012uea} in the $(m_{\tilde t},m_{LSP})$ plane.
  The stop is assumed to decay into jet and LSP with $100\%$  branching ratio everywhere.}
  \label{fig:razor}
\end{figure}
\begin{figure}[tp]
  \begin{center}
    \psfragfig*[width=0.7\textwidth]{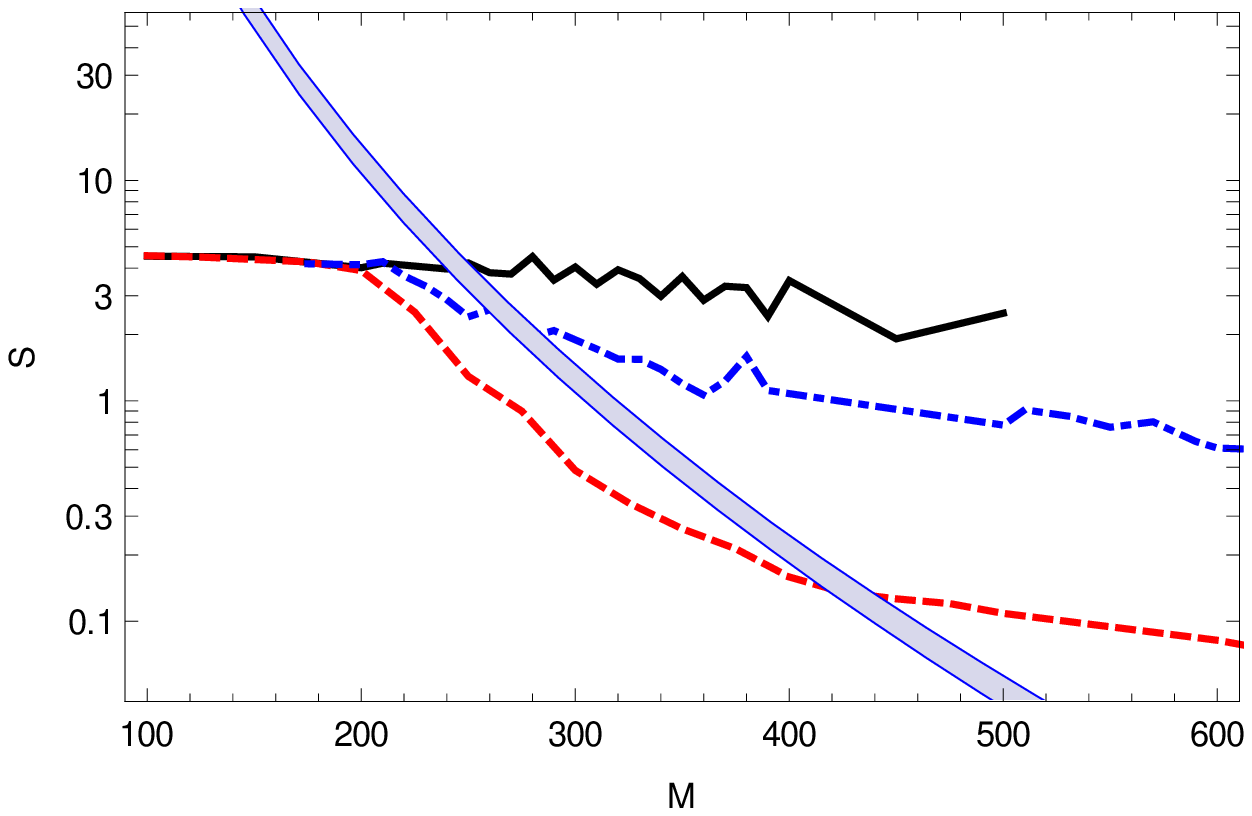}{
    \psfrag{M}[c][c]{$m_{\tilde{t}}$ [GeV]}
    \psfrag{S}[c][c]{$\sigma_{\tilde{t}^*\tilde{t}}$ [pb]}
    }
    \vspace{15mm}
    \\
    \psfragfig*[width=0.7\textwidth]{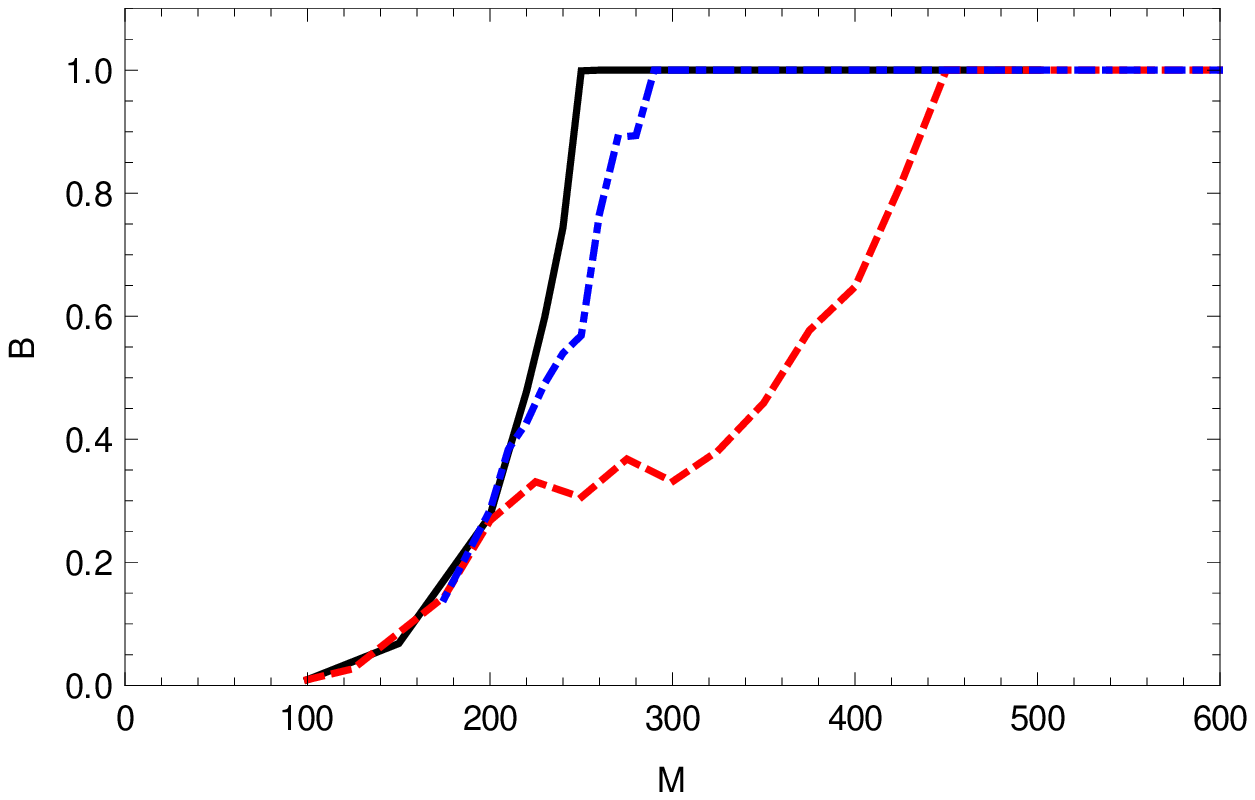}{
    \psfrag{M}[c][c]{$m_{\tilde{t}}$ [GeV]}
    \psfrag{B}[c][c]{Br$(\tilde{t}\to j+$LSP)}
    }
  \end{center}
  \caption{Cross section (top) and branching ratio (bottom)
  limits from the 7 TeV razor analysis as a function of stop mass for various LSP masses,
  (solid black $m_{\tilde{t}}-m_{\tilde{B}}=10$ GeV,
  dot-dashed blue: $m_{\tilde{t}}-m_{\tilde{B}}=150$ GeV,
  dashed red: $m_{\tilde{B}}=0$ GeV and $m_{3/2}=0$ GeV). 
  The gray band is the NLO stop production
  cross section at 7 TeV  computed using Prospino 2.1 \cite{Beenakker:1996ed}.}
  \label{fig:razor2}
\end{figure}
In figure  \ref{fig:razor} and figure  \ref{fig:razor2} we show the limits
obtained. 
Our results are consistent
with results in the literature
\cite{Delgado:2012eu,Dreiner:2012sh} to the extent they can be
compared.

\begin{figure}[tp]
  \begin{center}
    \includegraphics[width=0.7\textwidth]{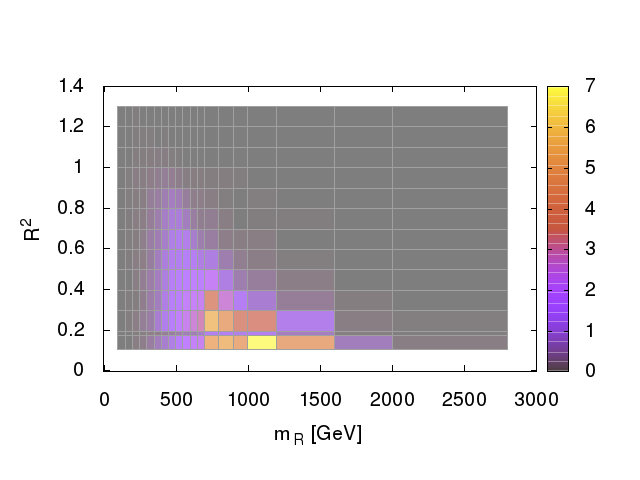}\\
    \includegraphics[width=0.7\textwidth]{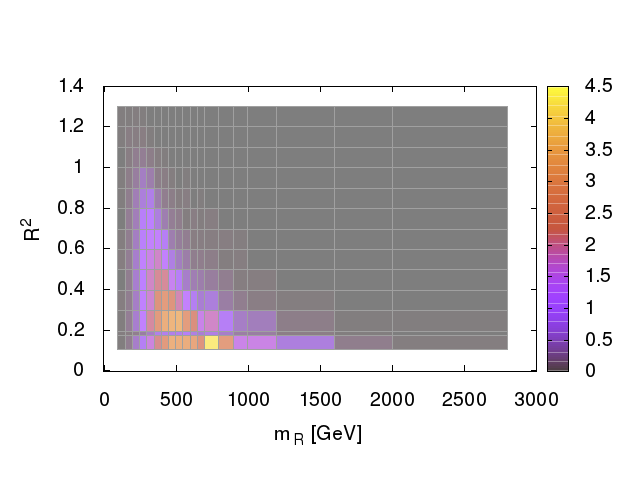}
  \end{center}
  \caption{Two-dimensional histograms for the number of events at $5$
  fb$^{-1}$ in $m_R$ and $R^2$ plane for 
  $m_{\tilde{t}} = 500$ GeV, $m_{\tilde B}=0$ GeV (top) and
  $m_{\tilde{t}} = 500$ GeV, $m_{\tilde B}=300$ GeV (bottom). }
  \label{fig:2drazor}
\end{figure}
We present the limits in the stop-LSP mass plane in figure \ref{fig:razor}, 
and the limits on the cross section and the branching ratio in figure
\ref{fig:razor2}. Similar
conclusions as the $\alpha_T$ search hold. The limit is insensitive to
the stop mass, and hence branching ratio suppression or multiplicity
enhancements do not change the limits significantly.

\paragraph{Extending the razor\\}

The razor analysis appears to be applicable beyond its design
topology.
This was already noted in previous analyses 
\cite{Fox:2012ee, Delgado:2012eu}. 
In particular, it is one of the strongest bounds on the
signatures considered in this paper. We comment briefly on the
possibility of extending the
current analysis strategy to improve the sensitivity
to our signal.

In figure \ref{fig:2drazor} we plot the two-dimensional
histogram of the number of
events expected at the LHC at $5$ fb$^{-1}$ as a function of the razor
variables, $m_R$ and $R^2$ for two mass spectra as examples. 
We see that for $m_{\tilde{t}} = 500$ GeV, $m_{\tilde{\tilde B}}=0$ GeV,
the events populate high $m_R$ bins, and hence would tend to pass the
analysis cuts. For $m_{\tilde{t}} = 500$ GeV,
$m_{\tilde{\tilde B}}=300$ GeV, the $m_R$ spectrum is significantly
softer. However, the events are pushed towards larger $R^2$ values,
where the QCD background is expected to be minimal. Therefore,
extending the razor signal region to include the low $m_R$ - high
$R^2$ region could potentially improve the sensitivity to the signal.

\section{Conclusions}

We have explored a scenario where the third generation squarks mix
significantly with either the first or the second generation. 
This arises generically 
(both in the left and the right-handed sector)
in models with an approximate R-symmetry, but 
at the same time, these results also apply to MSSM scenarios beyond
MFV where the
right handed stop mixes with the charm  (for instance
in the scenario described in  \cite{Blanke:2013uia}).
These frameworks are especially interesting in light of the recent 
 LHC null results for  degenerate squarks.
Also, independent of theoretical arguments, it is
important to make sure that the LHC has a good coverage of SUSY
scenarios beyond MSSM.

In this paper we have focused in
particular on a light stop somewhat degenerate with the LSP.
We study the FV decay topology, $\tilde t \to j$ LSP, which dominates
in an extended region of the parameter space for moderately large
mixing.
 This region
is unconstrained in the public analyses of ATLAS and CMS.  However,
our emulation of the CMS razor and $\alpha_T$ searches
\cite{CMS-PAS-SUS-11-003,Chatrchyan:2012uea} seem to be
sensitive to this region of the parameter space and are able to rule
out a significant portion of it. This is in agreement with recent work
on light stop phenomenology,
\cite{Krizka:2012ah,Delgado:2012eu,Dreiner:2012sh,Dreiner:2012gx}.
For a massive bino  we find that  the bounds on the stop mass varies between
$250$ GeV and 350 GeV depending on the mass splitting between the stop
and the bino. For a massless bino the bound  on FV decay mode rules
out all the stealth region $(m_{\tilde t}<250$ GeV)
 as long as mixing  is around $10 \%$.
For a massless gravitino LSP, a stop below 250 GeV is ruled by the FV
decay mode for an even smaller mixing, $3 \%$.
For heavier masses the dedicated searches  for $ \tilde t  \to t$ LSP
start to be efficient, setting a bound in the 500-560 GeV range
(depending on the mixing).

These light stop bounds involve extending searches in regions beyond
where they were designed to operate.  There is a potential danger that
this might not always  be possible.  For instance, the CMS razor
analysis \cite{CMS-PAS-SUS-12-009} attempting to put limits in the
mass region $ m_{\tilde q} <300$ GeV and $ m_{\tilde q}-m_{LSP} <150$
GeV finds that this region has a sensitivity to the ISR modeling in
simulation of signal events above a pre-defined tolerance; hence no
interpretation is presented for these model points.  This is an
important piece of information which draws attention to the challenges
for probing this region.  We encourage the experimental collaborations
to attempt to cover the light mass and/or somewhat compressed region
for all the searches which are potentially sensitive to it such as for
example the $ \alpha_T$ analysis and the razor analysis
\cite{CMS-PAS-SUS-11-003,Chatrchyan:2012uea}.

It is important to extend the LHC searches for light stops to all
possible channels unrestricted by theoretical bias. For instance, we
show that a light stop can have signatures similar to a first or
second generation squark in a significant region of parameter space.
This is especially relevant for the 7-8 TeV run of the LHC, since at
 13 TeV, the QCD backgrounds for jets+MET searches get even worse
    relative to the background for third generation final states. This
    could make our topology challenging to look for and potentially a
    blind spot for light stops.

\acknowledgments{ 
We thank Wolfgang Altmannshofer, Joe Lykken and Maurizio Pierini for
valuable discussions. We are grateful to Bogdan Dobrescu, Zackaria
Chacko, Marco Farina,Thomas Gr\`{e}goire, Roni Harnik and Felix Yu for
useful comments on the manuscript.
Fermilab is operated by Fermi Research Alliance, LLC  under Contract No. DE-AC02-07CH11359 with the United States Department of Energy.
}

\bibliography{squark-mixing}
\bibliographystyle{JHEP}

\end{document}